2. Author:  Robert L. Oldershaw

3. Address:  Physics Dept., Amherst College (Box 2262), Amherst, MA 01002 USA

   e-mail:  rlolders@unix.amherst.edu

10. Robert L. Oldershaw

**ABSTRACT:** After two decades of efforts to identify the enigmatic dark matter that comprises the dominant form of matter in our galaxy, the mass range for viable candidates appears to have been reduced by more than 50 orders of magnitude. Positive results have thus far been confined to the range: $10^{-7} M_o$ to $1.0 M_o$, with apparent clustering within the ranges $10^{-5} M_o$ to $10^{-3} M_o$ and $0.08 M_o$ to $0.5 M_o$. Positive and negative results are compared with specific predictions of cosmological models.

**Key Words:** Cosmology: dark matter, theory, miscellaneous, large-scale structure of Universe

**The Dark Matter Problem and Candidate Solutions**

  The dark matter problem arose during the 1930s when astronomers such as Zwicky, Oort and Kapteyn realized that the luminous and virial masses of galaxies differed by factors of 10 or more. Review papers on the history, theoretical aspects and empirical status of the dark matter problem have been published by Trimble (1987) and Carr (1994). In this paper we are concerned exclusively with observations of *galactic* dark matter, as opposed to intergalactic dark matter. This more limited subject was reviewed not long ago by Ashman (1992). The consensus that has emerged over the last two decades is that a mysterious non-luminous form of matter comprises most of a galaxy's mass, possibly as much as 90% to 99% of the total mass. The most likely location for this vast amount of non-luminous matter is thought to be an extensive galactic halo.

  The mass range for familiar dark matter candidates covers an incredible 78 orders of magnitude: from a putative $10^{-6}$ ev axion to $10^6$ $M_o$ black holes. In between there is an array of literally scores of candidate dark matter populations including 17 ev neutrinos,

"snowballs," quark "nuggets," primordial black holes and brown dwarfs (Trimble 1987). The empirical attack on the dark matter problem began in earnest in the 1970s, and several plausible candidates are thought to have been virtually ruled out, such as faint stars (Rieke 1989; Graff and Freese 1996a); gas, dust, rocks and snowballs (Hills 1986; Hegyi & Olive 1986); and massive ($> 1$ $M_o$) black holes (Bahcall et al. 1985).

Candidates that have survived the early rounds of falsification tests and remain the most viable possibilities are as follows. There is a rather large group of potential Cold Dark Matter (CDM) and Hot Dark Matter (HDM) candidates, with the leading ones being neutralinos (10 - 500 Gev), axions ($10^{-6}$ - $10^{-4}$ ev) and massive neutrinos (2 - 30 ev), as described by Dodelson et al. (1996). Brown dwarfs are a perennial favorite, though observations of truncated stellar mass functions may limit their potential contribution to the *total mass* of the dark matter (Williams et al. 1996; Graff and Freese 1996b). Current observational data permit white dwarf stars to remain viable candidates although this requires some assumptions that are difficult to defend (Adams & Laughlin 1996; Kawaler 1996). Low mass ($\leq 1$ $M_o$) neutron stars and primordial black holes are consistent with the available data, but scenarios for their formation remain sketchy (Trimble 1987; Carr 1994). Finally, there is the category of "other," which should certainly be included here since this category has had an extremely good track record throughout the history of science.

Because the field of galactic dark matter research is advancing rapidly, it is important to specify that the data and analyses presented here are those that have been published up to December 1996.

**Specific Predictions**

If a cosmological model is to have significant scientific value, then it must be able to retrodict a very large quantity of dark matter, and have something to say about its composition. Clearly a cosmological theory that is mute on the makeup of 50% to 99% of the universe is of limited utility. In this paper we concentrate on cosmological models that can retrodict the *galactic* dark matter in a natural way, and that can make *specific predictions* about the constituents of this universal and dominant form of matter.

The original Big Bang (BB) theory did not predict the existence of vast amounts of dark matter. However, when the BB model is amended by the Inflationary scenario (I), then large quantities of CDM and/or HDM are retrodicted, and a broad set of potential subatomic particle candidates can be identified (Dodelson et al. 1996). Technically this does not constitute a definitive prediction because the BB+I paradigm does not uniquely specify which of the many particle candidates are the major constituents. However, the BB+I models are currently the leading cosmological models, and the identification of a cluster of potential candidates is certainly a step in the right direction toward a definitive prediction. Therefore these models are included in the present discussion. As mentioned above, the axion, massive neutrino and neutralino are currently thought to be the best bet candidates of the standard BB+I models, although variations on the standard models lead to other "wimp" candidates or even to hydrogen in the form of cold molecular clouds (De Paolis et al. 1996)

The only cosmological model known to the present author that makes a definitive prediction about the galactic dark matter is a fractal model called the Self-Similar Cosmological Paradigm (Oldershaw 1987, 1989a, 1989b). Using its underlying principle of discrete cosmological self-similarity it was possible to predict in 1987 that the galactic dark matter *must* be dominated by two populations of ultracompact objects. The higher mass population was predicted to cluster tightly around a mass of 0.15 $M_o$ ($\pm$ 0.05 $M_o$), and would constitute about 90% of the galactic dark matter *mass*. The lower mass population would weigh in at $7 \times 10^{-5}$ $M_o$ ($\pm$ $2 \times 10^{-5}$ $M_o$) and roughly equal the larger mass population in terms of *numbers*. These quantitative predictions are truly definitive in that the model would be falsified if these predictions are not vindicated.

Apparently there are no further cosmological models that make specific predictions about the composition of the galactic dark matter. The Quasi-Steady State Cosmology (Hoyle et al. 1993) discusses different potential candidates with masses ranging over $\geq$ 16 orders of magnitude, and therefore does not meet the criteria for discussion in this paper. Theories that definitively predict that the galactic dark matter *does not exist* are assumed to be extremely unlikely and are not considered here.

**Observational Results**

Figure 1 shows the record of reported galactic dark matter mass estimates from the time that positive results first appeared in December 1991 until the present (December 1996). Table 1 contains references and quantitative information for these data points. The x-axis of Figure 1 depicts the full range of candidate galactic dark matter mass values as log M in units of $M_o$, and the y-axis is the time period from November 1991 to December 1996. Predicted mass values are shown as vertical structures: two discrete lines at -4.16 (=$7 \times 10^{-5}$ $M_o$) and -0.82 (= 0.15 $M_o$) for the fractal model predictions, and broad columns for the most likely potential ranges for the BB+I+HDM/CDM models: -

72.05 ($10^{-6}$ ev) to -70.05 ($10^{-4}$ ev), -65.75 (2 ev) to -64.57 (30 ev) and -56.05 (10 Gev) to -54.35 (500 Gev).

The data are presented in a temporal format for two basic reasons. Firstly, this type of presentation is effective in highlighting trends in the data. Secondly, in a case such as the galactic dark matter problem where analysis of the raw data is substantially model-dependent (Alcock et al. 1996; De Paolis et al. 1996b), there is the possibility of systematic errors in any analysis. By including several different analyses, involving different sets of data and assumptions, one is less likely to be misled by systematic errors. At this stage it is conceivable that an earlier estimate based on a small amount of data is closer to the actual value than a more recent and comprehensive result; in the long run the probability of this being the case should decrease.

When viewing Figure 1, two related features standout prominently: the positive results are confined to a relatively narrow segment ($10^{-7}$ $M_o$ to 1 $M_o$) of the full mass range, and they appear to form two clusters. The lack of positive results reported for the predicted BB+I+HDM/CDM ranges of $10^{-6}$ ev to 500 Gev is somewhat surprising. In spite of literally scores of clever and varied experimental designs, no reproducible candidate events have been reported. At one point there appeared to be some evidence for a 17 ev neutrino, but subsequent work has discredited this result (Schwarzschild 1993). There are numerous on-going searches for particle candidates, and many more are planned (Dodelson et al. 1996).

The substantial number of data points on the right side of Fig. 1 are all the products of gravitational microlensing experiments. The fundamental ideas of microlensing were discussed by Einstein (1936) and Refsdal (1964). The technical feasibility of using gravitational microlensing to attack the local dark matter problem was demonstrated in a key paper by Paczynski (1986). Subsequently, several research groups have taken up the challenge and two different observational approaches have been pursued. One approach focusses on the variability of macrolensed quasars, searching for brightening events that are best explained as microlensing events, as opposed to intrinsic variations. A second approach, taken up by the American/Australian MACHO group (Alcock et al. 1993), the European EROS group (Aubourg et al. 1993), the Princeton/Poland OGLE group (Udalski et al. 1993), and others, searches for local events wherein closer objects in our galaxy act as lenses for more distant galactic stars, or stars in neighboring galaxies.

Interestingly, each approach has yielded positive results that, with a few exceptions (to be discussed below), cluster in separate mass ranges. The quasar microlensing experiments have tended to yield evidence for dark matter objects in the $10^{-5.5}$ $M_o$ to $10^{-3}$ $M_o$ range, i.e., the planetary-mass range. These mass estimates are model-dependent and two are based on small sample sizes; therefore the margins of error are large, at least $\pm$ a factor of 10.

Local microlensing experiments, on the other hand, have found evidence for a large dark matter population residing in the halo and typified by masses on the order of 0.1 $M_o$. Two exceptions to this dichotomy are a possible finding of three planetary-mass events in EROS short-term variability data (Kerins 1995), and hints of stellar mass objects in quasar variability studies (Refsdal and Stabell 1993; Cummings and De Robertis 1995). The possibility that halo dark matter observations might be best explained in terms of a combination of planetary- and stellar-mass populations was suggested by Refsdal and Stabell (1993).

It should be noted that different authors present their mass results in different forms. Given a sample of estimated masses with a major peak at 0.1 $M_o$ and a smattering of larger masses, the average mass and the most probable mass can be *significantly different*. Also, different methods for calculating mass values yield different estimates. For example the MACHO group has recently reported (Pratt et al 1996) an *average* mass of about 0.5 $M_o$ for galactic dark matter objects in the halo, while the *most probable* mass would be significantly lower. Moreover, Jetzer (1996) analyzed essentially the same raw data with a method of moments analysis and came up with an average mass estimate that was lower by a factor of two, 0.27 $M_o$.

**Discussion**

This paper is intended as a brief overview of the dark matter problem and a progress report on efforts to actually identify the galactic dark matter objects. Figure 1 is a visual summary of the latter. For the time period covered here, the major implications of the empirical results are as follows.

> (1) There is a conspicuous absence of positive results reported for masses below $10^{-7}$ $M_o$, in sharp contrast to what is predicted by the BB+I paradigm. Persistent experimentation continues in the particle-mass range.

> (2) There is a conspicuous grouping of published positive results within the mass range $10^{-7}$ $M_o$ to 1.0 $M_o$. Also, the estimated masses of galactic dark matter objects appear to cluster tightly within the the mass range of 0.05 $M_o$ to 0.50 $M_o$, and more loosely within the mass range $10^{-7}$ $M_o$ to $10^{-3}$ $M_o$.

(3) The present results are consistent with the 1987 fractal model predictions of galactic dark matter mass peaks at 0.15 $M_o$ and $7 \times 10^{-5}$ $M_o$ (Oldershaw 1987, 1989a,b). Whether the MACHO, EROS, OGLE, etc. groups confirm a galactic dark matter peak in the predicted planetary-mass range is the next important test of the principle of discrete cosmological self-similarity. Events with durations of 0.5 to 1.0 days durations should be as numerous as those with 20 to 80 days, but more difficult to detect (Alcock et al 1996b). Previous negative results by the EROS and MACHO groups have been based on the assumption that <u>all</u> of the galactic dark matter is in the form of planetary-mass objects (Alcock et al. 1996b), whereas the fractal model predicts that the number of planetary-mass objects is *three orders of magnitude* lower than this assumption requires.

# TABLE 1

## GALACTIC DARK MATTER MASS ESTIMATES

| Seq. # | $\langle m \rangle$ ($M_o$) | $\langle m \rangle$ Log $M_o$ | Mon/Yr | Mon > 11/91 | Comments | Reference |
|---|---|---|---|---|---|---|
| 1 | $5.5 \times 10^{-5}$ | -4.26 | Dec 91 | 1 | Lensing event, component A, QSO 2237+0305 | Webster et al, 1991 |
| 2 | $10^{-5}$ | -5.00 | Oct 93 | 23 | QSO variability. Raises possibility of a planetary-mass + stellar-mass bimodal mass function | Refsdal and Stabell, 1993 |
| 3 | $10^{-4}$ | -4.00 | Nov 93 | 24 | $10^{-4}$ $M_o$ seems to give the best fit, but large uncertainty. Also see Hawkins, 1996 for comments | Schneider, 1993 |
| 4 | $10^{-7}$ | -7.00 | Oct 95 | 47 | Reanalysis of EROS short-term data suggests possibility of several planetary-mass events | Kerins, 1995 |
| 5 | $10^{-3}$ | -3.00 | Feb 96 | 51 | QSO variability | Hawkins, 1996 |
| 6 | $10^{-5}$ | -5.00 | Jun 96 | 55 | QSO variability - strong peak in planetary-mass range | Schild, 1996 |
| 7 | $10^{-5.5}$ | -5.50 | Sep 96 | 58 | Second analysis of data in #5 | Schild and Thompson, 1996 |
| 8 | 0.12 | -0.92 | Oct 93 | 23 | 1st MACHO event (halo) | Alcock, et al, 1993 |
| 9 | 0.2 | -0.70 | Oct 93 | 23 | EROS #1 and #2 (halo) | Aubourg, et al, 1993 |
| 10 | 0.144 | -0.84 | Mar 94 | 28 | Methods of momments analysis of MACHO #1 and EROS #1 + #2 | Jetzer and Masso, 1994 |
| 11 | 0.08 | -1.10 | Apr 94 | 29 | MACHO #1 and EROS #1 + #2 | Evans, 1994 |
| 12 | 0.08 | -1.10 | Sep 94 | 34 | Method of moments analalysis of MACHO #1-#3 and EROS #1 + #2 | Jetzer, 1994 |
| 13 | 0.08 | -1.10 | Apr 96 | 53 | MACHO #1-#3 | Alcock, et al, 1996a |
| 14 | 0.27 | -0.57 | May 96 | 54 | MACHO #1-#8 and EROS #1 + #2 | Jetzer, 1996 |
| 15 | 0.50 | -0.30 | Jun 96 | 55 | MACHO #1-#8 | Pratt, et al, 1996 |
| 16 | 0.40 | -0.40 | Aug 96 | 57 | MACHO #1-#7 and EROS #1 + #2 | Flynn, et al, 1996 |

**Figure 1**

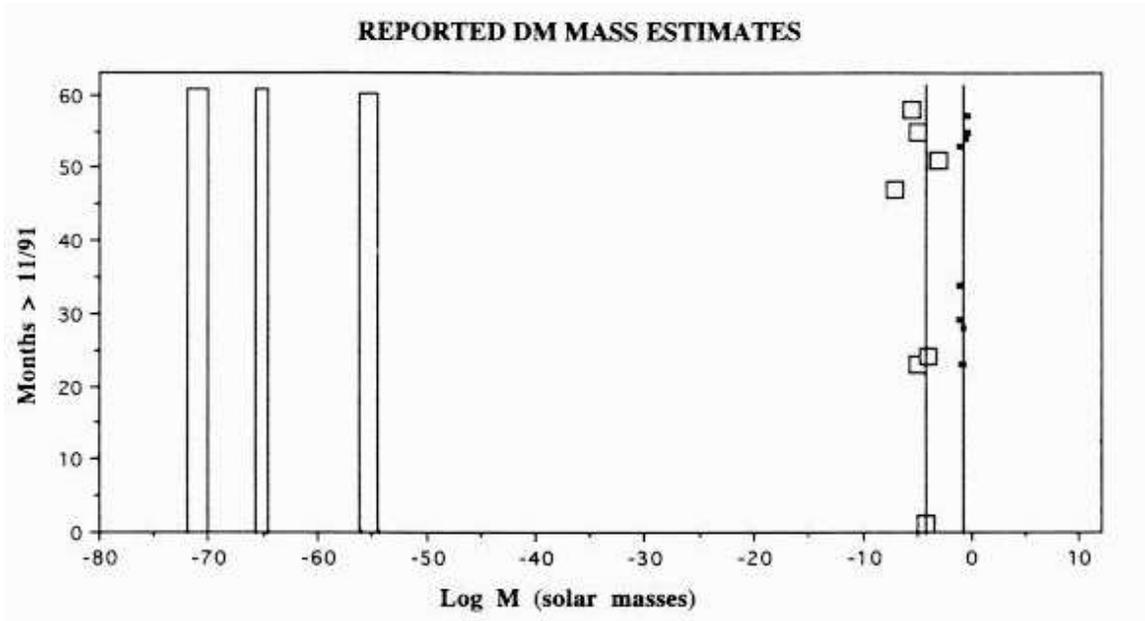